\documentstyle[mymulticol,pra,aps,amsfonts]{revtex}

\newcommand{\cX}{{\cal X}}
\newcommand{\cY}{{\cal Y}}

\newcommand{\beq}{\begin{equation}}
\newcommand{\eeq}{\end{equation}}
\newcommand{\beqy}{\begin{eqnarray}}
\newcommand{\eeqy}{\end{eqnarray}}

\newtheorem{Definition}{Definition}

\newtheorem{Theorem}{Theorem}

\newenvironment{Proof}{{\it Proof: \,}}{$\Box$ \vspace{0.3cm}}

\newenvironment{Definition*}{{\bf Definition}}{}

\makeatletter
\def\@beginTheorem#1#2{\trivlist \item[\hskip \labelsep{\bf #1\ #2}]}
\def\@opargbegintheorem#1#2#3{ \trivlist
      \item[\hskip \labelsep{\bf #1\ #2\ (#3)}]}
\makeatother

\makeatletter
\def\@beginLemma#1#2{\trivlist \item[\hskip \labelsep{\bf #1\ #2}]}
\def\@opargbeginLemma#1#2#3{ \trivlist
      \item[\hski
 Hence we have the same statements
about the increase of the supports for increasing depth, where
the local transformations are not counted for the depth.
p \labelsep{\bf #1\ #2\ (#3)}]}
\makeatother

\makeatletter
\def\@beginDefinition#1#2{\trivlist \item[\hskip \labelsep{\bf #1\ #2}]}
\def\@opargbeginDefinition#1#2#3{ \trivlist
      \item[\hskip \labelsep{\bf #1\ #2\ (#3)}]}
\makeatother

\makeatletter
\def\@beginCorollary#1#2{\trivlist \item[\hskip \labelsep{\bf #1\ #2}]}
\def\@opargbeginCorollary#1#2#3{ \trivlist
      \item[\hskip \labelsep{\bf #1\ #2\ (#3)}]}
\makeatother

\makeatletter
\def\@beginExample#1#2{\trivlist \item[\hskip \labelsep{\bf #1\ #2}]}
\def\@opargbeginExample#1#2#3{ \trivlist
      \item[\hskip \labelsep{\bf #1\ #2\ (#3)}]}
\makeatother

\def\C{{\mathbb{C}}}

\def\R{{\mathbb{R}}}

\def\N{{\mathbb{N}}}

\newcommand{\cH}{{\cal H}}

\newcommand{\cS}{{\cal S}}

\title{A complexity measure 
 for continuous time quantum algorithms}

\author{D. Janzing\thanks{Electronic address: janzing@ira.uka.de}  and 
Th. Beth}
\address{Institut f\"ur Algorithmen und Kognitive Systeme, Am Fasanengarten 3a,
    D--76\,131 Karlsruhe, Germany}

\begin{document}
\maketitle

\begin{abstract}
We consider unitary dynamical evolutions on $n$
 qubits caused by time 
dependent pair-interaction Hamiltonians and 
show that the running time of a parallelized two-qubit gate network 
simulating the evolution is given by the time integral over the
chromatic index of the interaction graph.
This defines a
complexity measure of continuous and discrete quantum algorithms 
which are  in
exact one-to-one  correspondence. Furthermore we prove a lower bound 
on the growth of large-scale entanglement depending on the
chromatic index.
\end{abstract}

\begin{multicols}{2}

\section{Introduction}
At the moment, the most popular model of a quantum computer
consists of the  $2^n$ dimensional Hilbert space 
$\cH_n:= (\C^2)^{\otimes n}$ of
 `$n$ qubits' as its memory space, and some  one- 
and two-qubit gates as its set of basic 
transformations (see e.g. \cite{Ah}).
There are several reasons for taking one- and two-qubit gates as the basic 
ones:
Firstly, from the pure mathematical point of view, it is  quite
natural to look for a subset of the Lie group of unitary transformations on
the Hilbert space $\cH_n$  generating the whole group.
Obviously, one-qubit operations do not generate the whole Lie group,
the set of two-qubit gates does \cite{Di}. Hence there is no reason for 
 taking  more complicated
 transformations like three-qubit unitary operators as basic ones.
Secondly, the model of two-qubit gates might be considered as the attempt
to develop quantum computation in strong analogy to the theory of
classical devices:
Building complex logical networks from two-bit gates is a successful
concept of classical computation.
The third reason stems from physics. Unfortunately it
 is  only `a little bit' convincing:
 From the fundamental point of view, $n$ particles
interact always in the form of pair-interactions, i.e., the total Hamiltonian
$H$ of the system can be decomposed as 
\[
H=\sum_{k,l\leq n}H_{k,l}+\sum_{j\leq n} H_j
\]
 where $H_{k,l}$ is a self-adjoint operator acting on 
the joint Hilbert space of  particle $k$ and $l$ and $H_j$ is the free
Hamiltonian of particle $j$ (without loss of generality, we can drop
the second sum, by reckoning it to the first part).
Pair-interactions $H_{k,l}$ are the infinitesimal versions of two-qubit gates:
every unitary of the form $e^{i H_{k,l}t}$ for $t\in \R$ 
is a two-qubit gate.
On the one hand, this seems to be an important justification for two-qubit 
gates, since it refers to the form of the fundamental forces of nature,
 on the other hand, the argument is not really correct:
In general there is no obvious correspondence
between  the time evolution 
\begin{equation}\label{stetig}
e^{i\sum H_{kl}t}
\end{equation}
and  any finite  sequence of two-qubit gates. However,
there is an obvious
 simulation by two-qubit gates  in an {\it approximative} sense
given by the well-known Trotter formula:
\begin{equation}\label{Trotter}
\lim_{m\to\infty}(\Pi_{k,l} e^{i H_{k,l}t/m})^m=e^{i\sum_{k,l}H_{k,l}t}.
\end{equation}
This example shows, that the simulation of the time evolution
caused by a time-independent pair-interaction Hamiltonian
might require an {\it infinite} number of two qubit gates.
Hence it seems to
 suggest that a definition of complexity based on
two-qubit gates does {\it not} take into account the most
natural form of dynamics of many-particle quantum physics.
However,
it has been shown \cite{Ll} that the number $m$
 of gates required to simulate the right hand side of (\ref{Trotter})
up to an error $\epsilon$ is only  growing
with $t^2/\epsilon$.
Despite the fact that infinite accuracy requires an infinite number
of gates, the {\it time} for implementing the growing number
of unitaries  does not tend to infinity if one assumes that the implementation
of $e^{iH_{k,l}t/m}$ requires the time $O(t/m)$ (see \cite{Ll}).
Taking this assumption,
we will show that the running time of a discrete quantum algorithm
for simulating the dynamics given by (\ref{stetig}) is determined by
 the chromatic index of the interaction graph.
This turns out to be true even for time-dependent pair-interactions.
One might reformulate this result by saying that the running time depends
on `the complexity of the interaction'.
The relevance of this result is twofold:
In case realizations of future quantum computers are based on
two-qubit gates, it gives exact statements about the complexity of simulating 
non-autonomous quantum dynamics resulting from pair-interaction Hamiltonians.
To our knowledge, this is the first exact analogue between 
complexity measures of discrete and continuous quantum algorithms.
Secondly, our result is relevant for the simulation  
of arbitrary pair-interaction dynamics by
a dynamic with restricted interaction graph: We derive statements
about the efficiency for simulating time  evolutions based on pair-interactions
with {\it high chromatic index} by other evolutions with  interactions 
of {\it lower 
index}.

\section{Discrete and continuous quantum algorithms} 

In our model of discrete quantum computers,
 two-qubit gates  acting on disjoint pairs of qubits
can be implemented simultaneously. We define:

\begin{Definition}
A {\bf discrete quantum algorithm}  $A$ {\bf of depth} $k$ is a sequence of
$k$ steps $\{A_1,\dots,A_k\}$ where every step consists of a set
of two qubit gates  $\{u_{jl}\}_{j,l}$ acting on disjoint
pairs $(j,l)$ of qubits.
Every step $i$ defines a unitary operator $v_i$ by taking the product
of all corresponding unitaries in any order.
The product $u:=\Pi_{i\leq k} v_i$ is the `unitary operator implemented
by $A$'. 
\end{Definition}

The following quantity measures the deviation of a unitary operator
from the identity:

\begin{Definition}
The {\bf angle} of an arbitrary unitary operator $u$
 is the smallest possible norm\footnote{Here and in the following $\|.\|$ denotes the operator norm given by $\|a\|:=\max_x \|ax\|$ where $x$ runs over the unit vectors of the corresponding Hilbert space.} $\|a\|$ of a selfadjoint
 operator $a$ which satisfies
$e^{ia}=u$.
It coincides with the time required for the implementation of $u$ if
the norm of the used Hamiltonian is 1.
\end{Definition}

This term allows us to formulate a modification of the term `depth'
which will later turn out to be decisive in connecting complexity measures
of discrete and continuous algorithms:

\begin{Definition}
Let $\alpha_i$ be the maximal angle of the unitaries performed
in step $i$. Then the {\bf weighted depth} is defined to be
the sum $\sum_i\alpha_i$.
\end{Definition}

 Assuming that the implementation time
of a unitary is proportional to its angle, the weighted depth is the 
running time of the algorithm.
Since  this  coincidence is based on a
possibly unrealistic assumption we will prefer the term 
`weighted depth'
in formulating exact mathematical statements.

Now we want to formalize the notion of quantum  algorithms
based on time dependent Hamiltonians. Such
continuous algorithms have already been  considered in the 
literature\footnote{In \cite{FG} time-dependent Hamiltonian algorithms are called an
`analog analogue of a digital quantum computer'.}
(e.g. see \cite{FG,ZR,PZ,FKW,ZZF,MDA}),
 but in our approach the Hamiltonians are 
explicitly restricted to pair-interactions in many-particle systems:

\begin{Definition}\label{calg}
A {\bf continuous quantum algorithm $A$ of running time} $T$ 
is a piecewise Lipschitz continuous function
$t\mapsto H(t) $ from the interval $[0,T]$  into the set of those
 self-adjoint
operators acting on $\cH_n$ which can be decomposed into
\[
H(t)=\sum_{k,l}H_{k,l}(t),
\]
where every $H_{k,l}$ is a self-adjoint operator acting on the qubits
 $k$ and $l$.
Here $H_{k,l}=H_{l,k}$ and the sum runs over the unordered pairs.
We say `$A$ implements $u$' if $u=u_T$ and $(u_t)_{t\in T}$
 is the solution of the non-autonomous 
differential equation $\frac{d}{dt}u_t=-i H(t)$ with $u_0=1$. 
\end{Definition}

As already mentioned above,  the weighted depth of a discrete
algorithm
 for simulating the evolution (\ref{stetig}) is finite. 
More precisely, it is at most
 $mt$ if $m$ is the number
of pairs $(k,l)$ with the property that $H_{k,l}\neq 0$. 
But, if one has additional information about the interactions, one 
can make further statements about possible parallelization. This is illustrated
by the following examples:
In case all the pairs $(k,l)$  with $H_{k,l}\neq 0$ are disjoint,
we can perform all the unitaries $e^{iH_{k,l}t/m}$ simultaneously
and get the running time $t$.
In case of the nearest-neighbor interaction in a one dimensional 
spin chain, we have only pairs of the form $(k,k+1)$. Hence we can perform
all the transformations 
\[
e^{i H_{2k,2k+1}t/m}
\]
 simultaneously and 
the operations 
\[
e^{iH_{2k+1,2k+2}t/m}
\]
 as well.
Here, parallelization allows to decrease the running time down to $2t$.

We show that the graph theoretical concept 
of chromatic index  offers the appropriate terminology
for determining the degree of  possible parallelization.

\begin{Definition}\cite{Bo}\label{chr}
The {\bf chromatic index} of a graph with vertices $\{1,\dots,n\}$ is
the least number of colors required for coloring
the edges in such a way that there are no edges with the same color
having a common vertex.
\end{Definition}

For every time $t$ during the running time of the continuous algorithm
we define a family of undirected  graphs:

\begin{Definition}
For every non-negative  real number $r$ and every time $t\in [0,T]$ we
define the {\bf interaction graph} $G_r(t)$ as follows: 
Take the qubits $\{1,\dots,n\}$ as vertices and let the  edges
be all the pairs
$(k,l)$ with the property $\|H_{k,l}(t)\|> r$.
\end{Definition}

The following quantity turns out to be decisive for the degree
of possible parallelization of the discrete simulation:

\begin{Definition}
Let $n_r(t)$ be the chromatic index of  $G_r(t)$.  
For any time $t\in [0,T]$ of  a continuous  algorithm
we define the {\bf weighted chromatic index} $W(t)$ as
\[
W(t):=\int_0^\infty n_r(t)dr.
\]
\end{Definition}

Furthermore we will need the time integral of $W$:

\begin{Definition}

The {\bf integrated chromatic index} of the quantum algorithm
is defined to be 
\[
I:=\int_0^T\int_0^\infty n_r(t)drdt,
\]
where $T$ is the running time of the continuous algorithm.
\end{Definition}

Actually, the terminology `integrated weighted chromatic index'
would be appropriate. We preferred the short terminology.

\section{Translation between discrete and continuous algorithms} 

The following theorem suggests that the simulation of an interaction with high
integrated chromatic index generically requires  a discrete algorithm
with high weighted depth. More precisely, it shows
that the complexity of a unitary transformation can
equivalently be  defined as the infimum over the values of the weighted depth
of all possible discrete implementations or the infimum over the values 
of the integrated chromatic index of all possible continuous implementations:

\begin{Theorem}\label{Haupt}
\hspace{0cm}

\noindent
a)
Every 
 arbitrary unitary operator $u$ acting on n qubits 
which can be implemented by 
a discrete quantum algorithm 
with weighted depth $\alpha$ can also be implemented by
a continuous quantum algorithm with integrated chromatic index $\alpha$.

\noindent
b)
If there is a continuous algorithm with integrated chromatic index $\alpha$ 
implementing $u$ then there is a sequence of discrete algorithms $(A_k)_{k\in \N}$
 implementing the 
unitaries $u_k$  such that $\lim_{k\to\infty}
 u_k=u$ and the corresponding values of the  weighted depth
 converge to $\alpha$.
\end{Theorem}

\begin{Proof}
Statement a) is almost trivial:
Let $A$ be a discrete quantum algorithm of depth $k$ with the maximal angles
$(\alpha_j)_{j\leq k}$. Let $\cS_j$ be the set of qubit pairs which are 
addressed in the step $j$ and $(u_p)_{p\in \cS_j}$ the corresponding
set of unitary transformations. Define
a continuous algorithm with running time $k$ as follows:
for $t\in (j-1,j)$ we\footnote{Here we denote an interval 
by (.) if we do not care about
whether it is open or closed.}
  define the constant Hamiltonian $H(t):=\sum_{p \in \cS_j} -i\ln u_{p}$.
Since the angle of a unitary operator $u$ is the norm of $\ln u$,
we have $\alpha_j=\max_{p\in \cS_j}\|\ln u_{p}\|$. Hence,
for every $t$ in the interval $(j-1,j)$ we have $n_r(t)=0$ for 
$r\geq \alpha_j$ and $n_r(t) =1$ otherwise. 
Obviously we have 
\[
\int_0^T\int_0^\infty n_r(t)drdt= \sum_j \alpha_j.
\]

For statement b)
take a partition of $[0,T]$ into small  intervals
 on which 
$H$ is continuous. 
Let $(t,t+\epsilon)$ be one of those intervals. 
Let 
\[
r_1<\dots <r_m
\]
be the positive numbers $(\|H_{i,j}(t+\epsilon/2)\|)_{i,j\leq n}$ in its
canonical ordering. Furthermore define $r_0:=0$. 
Obviously, the function $r\mapsto n_r(t+\epsilon/2)$
takes  constant values $n_j$  on every interval $]r_{j-1},r_{j}]$ for
 $1\leq j\leq m$.
Let 
\[
M_1\cup M_2\cup \dots \cup M_{n_j}
\]
 be a partition of the set of 
 edges 
of $G_{r_j}(t+\epsilon/2)$ given by an
 allowed coloring (in the sense of Definition 
\ref{chr}) of the
 edges corresponding
to the chromatic index $n_j$.
For every $j\leq m$ we proceed as follows:
For every $M_p$ define a step of a discrete algorithm by
the set of unitaries 
\[
\Big\{e^{(i\epsilon H_{k,l}(t+\epsilon/2)) \times (r_{j}-r_{j-1})/\|H_{k,l}
(t+\epsilon/2)\|}\Big\}_{(k,l) \in M_p}.
\]
Since $p$ runs from $1$ to $n_j$ we obtain $n_j$ steps for the discrete
algorithm. Every step has at most the angle  $\epsilon (r_{j}-r_{j-1})$.
 The weighted depth of these 
$n_j$ steps is at most  $n_j \epsilon (r_{j}-r_{j-1})$. 
Doing this for every $j\leq m$, we obtain a
 sequence of $\sum_{j\leq m} n_j$ steps which substitutes 
the continuous algorithm
on the interval $(t,t+\epsilon)$ up to an error in the  order of  $\epsilon^2$.
The weighted depth of this sequence is at most   
\begin{equation}\label{weighted}
\sum_{1\leq j\leq m} n_j \epsilon(r_{j}-r_{j-1})=\epsilon 
\int_0^\infty n_r(t+\epsilon/2)dr.
\end{equation}
Without loss of generality we will assume the weighted depth
to be {\it equal} to the
 terms in equation (\ref{weighted}) since we can blow up the
algorithm by transformations which cancel each other out.
If the total discrete algorithm is defined by
combining the sequences for every interval of the form $(s,s+\epsilon)$
we get a simulation of the continuous algorithm 
with  an error in the order of $\epsilon$. Furthermore
 the total weighted depth converges to $\int_0^T\int_0^\infty n_r(t)dt$
for $\epsilon\rightarrow 0$.
\end{Proof}

One can consider discrete quantum algorithms as special cases of continuous
quantum algorithms in the sense of the proof of part a) of Theorem \ref{Haupt}.
Then the discrete quantum computer is obtained by restricting the
 Hamiltonians $H(t)$ to those with chromatic index one and
norm one for each non-vanishing component $H_{k,l}(t)$. 
In this sense, part a) of Theorem \ref{Haupt} is a special case 
of the  general principle
that continuous time algorithms 
can be simulated by other continuous time algorithms with lower chromatic 
indices and the same strength of  the pair-interactions
 where the running time is increased by the quotient
of the chromatic indices. 
We sketch this simple observation in a not too formal way:

At time  $t$ we have the 
interaction 
\[
H(t):=\sum_{(j,l)} H_{j,l}(t),
\]
where $(j,l)$  runs over the edges $E$ of the graph $G_0(t)$.
For $m< n_0(t)$ take $k$ such that $km\geq n_0(t)$.
Take a partition of the edges of $G_0(t)$
 into $k$ subsets such that
the  subgraphs  with the set of edges $(E_i)_{i\leq k}$ 
 have chromatic indices
less or equal to $m$. 
Substitute the small time interval
$(t,t+\epsilon)$ of the original algorithm by the
$k$ time intervals $(t+(i-1)\epsilon,t+i\epsilon)$
with $1\leq i\leq k$ in which
the Hamiltonian $\sum_{(j,l)\in E_i} H_{j,l}(t+\epsilon/2)$
is switched on. The difference of the unitary operators implemented by the
 substituted algorithm and the original one is of second order in $\epsilon$
for every subinterval. Hence the total error
tends to zero for $\epsilon \rightarrow 0$.

This observation justifies in some sense our point of view 
that one might think of the chromatic index as the complexity of
the interaction. This raises the question whether bounds
can be given on the required integrated chromatic index necessary
for preparing certain entangled quantum states from an initial product state.

\section{Complexity bounds for high entanglement}

In \cite{JB1} we gave lower bounds on the
depth required for  preparing   certain  states with large-scale entanglement 
(some of those bounds are easy conclusions from \cite{AKN}) for large qubit numbers $n$. 
More specifically, one can show the following: 
Let $(a_j)_{j\leq n}$ be a family of  arbitrary self-adjoint
 operators  where every $a_j$ acts on qubit $j$ and has operator norm 1.
For product states, the   
variance of the observable 
\begin{equation}\label{meanfield}
a:=\sum_{j\leq n} a_j
\end{equation}
 grows with $O(n)$. Values of
the variance in the order of $n^2$
indicate large-scale entanglement for pure states \cite{JB2}.
  In a state obtained by
a discrete algorithm with depth $k$ we could show \cite{JB1} 
the variance to be less
or equal to $n\,4^k/2$. Hence the emergence of large-scale entanglement
in the sense described above requires a depth in the order of $\log n$. 

It is natural to ask whether similar bounds can be shown for 
the {\it weighted} depth and the integrated chromatic index.
However,
below we present a proof for such a bound. In contrast to the discrete 
definition of depth the bound
is asymptotically independent on $n$ in the limit $n\to\infty$.
Whether this is a lack of our estimations or whether it is a hint for
 a fundamental difference between complexity measures
 of continuous and discrete
algorithms is unclear.
We have:

\begin{Theorem}\label{entan}
 Let $t\mapsto H(t)$ be a continuous quantum algorithm with
integrated chromatic index $\alpha$ implementing the unitary $u$. Let
the quantum computer start 
in the product state $\rho^{\otimes n}$. 
Let $\sigma:=u(\rho^{\otimes n})u^\dagger$  be the state obtained by
the algorithm.
Then for the variance of any observable $a$ of the form  (\ref{meanfield}) 
in the state $\sigma$ we have
\[
V_\sigma (a)\leq \frac{n}{(1-2\alpha)^4}
\]
with
\begin{eqnarray*}
V_\sigma(a)&:=&tr(\sigma a^2)-
(tr(\sigma a))^2\\&=&tr(\rho^{\otimes n}(u^\dagger au)( u^\dagger au))
-tr(\rho^{\otimes n}(u^\dagger au))^2 .
\end{eqnarray*}
\end{Theorem}

\begin{Proof}
Due to Theorem \ref{Haupt} 
we can restrict the proof to the case that 
 $H$ is of the same form as that one constructed 
in the proof of Theorem \ref{Haupt} a), i.e.,
$H(t)$ is for every $t$ a sum of pair-interactions acting on
 disjoint pairs. Furthermore we will assume without loss of 
generality that the norm of every non-vanishing pair-interaction
is one.
Then  $\alpha$ coincides with the running time
$T$.
We write $H(t)=\sum_{e\in E(t)} H_e(t)$ where
 $E(t)$ is the set of edges
of the interaction graph $G_0(t)$.
By solving the differential equation 
\[
\frac{d}{dt}a(t)=i[H(t),a(t)]
\]
with $a(0)=a$ 
we obtain $u^\dagger au=a(T)$. Explicitly, $u^\dagger a u$ is given by
a Dyson series \cite{Ze}:
\[
u^\dagger au=\sum_{k\in \N} i^k
\int 
[H(t_k),[\dots,  [H(t_1),a]\dots]] dt^k,
\]
where the integration is carried out over the simplex
\[
0\leq t_1\leq \dots \leq t_k\leq T.
\]

Let $X:=(p_k,\dots, p_1,r)$ be a $k+1$ tuple consisting of
$k$ unordered pairs  $p_j$ 
(which are considered as subsets of $\{1,\dots,n\}$)
 and one element $r\in \{1,\dots,n\}$.
Define
\begin{eqnarray*}
&&B_X(t_k,\dots,t_1):=\\&&[H_{p_k}(t_k),[H_{p_{k-1}}(t_{k-1}),[
\dots,[H_{p_1}(t_1),a_r]\dots]].
\end{eqnarray*}
Furthermore, for an arbitrary pair of observables $c,d$ we 
introduce the covariance $C(c,d)$ in the state $\rho^{\otimes n}$
as 
\[
C(c,d):=tr(\rho^{\otimes n}cd)-tr(\rho^{\otimes n} c)
tr(\rho^{\otimes n}d).
\]

Then the variance of the observable $a$ in the state $\sigma$ can be 
written as: 
\[
V_\sigma (a)=\sum_{k,l} F_{k,l}
\]
with
\[
F_{k,l}:=\int \int
\sum_{X,Y}  C (B_X(t_k,\dots,t_1)B_Y(s_l,\dots, s_1))dt^kds^l,
\]
where the sum runs over all
$(k+1)$-tuples $X$ of the form
$X:=(e_k,\dots,e_1,r)$
with $e_j\in E(t_j)\,\forall j$ 
and $r \in \{1,\dots,n\}$ and $(l+1)$-tuples $Y$ which are defined analogously 
by using the sets of edges
$E(s_j)$ instead of $E(t_j)$ (here $k$ runs over all non-negative integers).
In the following, we shall sometimes consider an $X$ of this form
canonically as a subset of $\{1,\dots,n\}$ and write
$j\in X$ iff $j$ agrees with the rightmost element of $X$ {\it or} 
with one of the vertices
of at least one $e_i$ for $1\leq i \leq k$.

Now let the times $t_k,\dots,t_1$ and $s_l,\dots,s_1$ be fixed.
We define $\cX_k$ as the set of $(k+1)$-tuples
$(e_k,\dots,e_1,r)$ with 
$e_j\in E(t_j)$   and
the property that every
$e_j$ has a vertex which is an element of $(e_{j-1},\dots,e_1,r)$.
It is easy to see that $B_X(t_k,\dots,t_1)\neq 0$
 implies $X\in \cX_k$. 
 The set $\cX_k$ can be constructed iteratively:
Set $\cX_0:=\{(1),\dots,(n)\}$.
Then we have $(e_k,X)\in \cX_k$ if and only if $X\in \cX_{k-1}$ and
$e_k \in E(t_k)$  and at least one of the  vertices of $e_k$
is an element of $X$ (here $(e_k,X)$ is an informal notation
for  the $(k+1)$-tuple obtained by appending $e_k$ to the $k$-tuple $X$).
In an analogous way we define the set $\cY_l$ by referring to the sets
$E(s_j)$ instead of $E(t_j)$.

The covariance
\begin{equation}\label{Cov}
C(B_X(t_k,\dots,t_1)B_Y(s_l,\dots,s_1))
\end{equation}
can only be nonzero if $X\in \cX_k$ and $Y\in \cY_l$
 have nonempty intersection
(considered as subsets of $\{1,\dots,n\}$).
We show that there are at most 
\newline
$n (k+1)!(l+1)!$ pairs
$X,Y$ with this property by proving that
for every $j\in \{1,\dots,n\}$ there are at most $(k+1)!$ 
sets $X\in \cX_k$ with $j\in X$. This can be seen by
 induction over $k$:
For $k=0$ the statement is obvious.
Assume the statement to be true for $\cX_{k-1}$.
For every $X\in \cX_{k-1}$
 with $j\in X$ we have at most $k+1$ elements
of the form $(e_k,X)\in \cX_k$ since for each vertex  $v\in X$
there can be at most one $e_k\in E(t_k)$ such that 
$e_k$ has $v$ as a vertex (remember that the corresponding
graph $G_0(t_k)$ has chromatic index 1). 
If $j\not\in X$ the statement $j\in (e_k,X)$ can only be true
if $j\in e_k\in E(t_k)$. There is at most one $e_k\in E(t_k)$ 
with vertex 
$j$. Let $e_k$ be the pair $\{j,m\}$. By assumption there are at most
$k!$ tuples $X\in \cX_{k-1}$ such that $m\in X$. Hence there are
at most $(k+1)k!$  sets $(e_k,X)\in \cX_k$ with $j\in (e_k,X)$.

Since the norm of every $H_e(t)$ is zero or one by assumption,
the norm of $B_X(t_k,\dots,t_1)$ can never exceed $2^k$.

Hence we get
\begin{eqnarray*}
&\sum_{X\in \cX_k,Y\in \cY_l}& C(B_X(t_k,\dots,t_1),B_Y(s_l,\dots,s_1)) \\
&\leq & n\, 2^k\,2^l\,(k+1)!\,(l+1)!.
\end{eqnarray*}
 Since the integration is carried out over a
 simplex of size $T^kT^l/(k!\,l!)$
we obtain:
\begin{eqnarray*}
V_\sigma (a)&\leq& n\sum_{k,l} (2T)^k(k+1) (2T)^l(l+1)\\&=&
n \big(\sum_k (2T)^k(k+1)\big)^2.
\end{eqnarray*}
This power series converges for $2T<1$. 
Since 
\[
\sum_{k\in \N} (k+1)p^k=1/(1-p)^2 \hbox{  for every } -1<p<1
\]
we have:
\[
V_\sigma (a)\leq \frac{n}{(1-2T)^4}.
\]
\end{Proof} 

Theorem \ref{entan} shows that for huge $n$ 
the variance of the measurement results
of any  mean-field observable can only grow in the order of $n^2$
if the integrated chromatic index $\alpha$  is at least $1/2$.
Hence $\alpha\geq 1/2$ is a necessary condition 
for the particular class of large-scale entangled states
considered in  \cite{JB2}.

\section{Conclusions}

We have shown that the running time of a discrete algorithm
simulating the evolution defined by a time-dependent 
Schr\"{o}dinger-equation with pair-interaction Hamiltonians
is given by the  time integral over the chromatic index
 of the interaction
 graph. This result suggests to take this time integral as a 
complexity measure for continuous time algorithms
and it seems natural to ask for the complexity for the preparation
of highly entangled states in many-particle systems.
We proved such a bound.
At the moment, we cannot decide
whether it is tight or not. 
Given a quantum system with the property that the weighted chromatic index
$W(t)$ of the relevant interaction Hamiltonian 
satisfies
$W(t)\leq m$ for every time $t$ we conclude that
the system requires at least the time $1/(2m)$ in order
 to produce large-scale entanglement in the sense explained above.
If we assume a non-vanishing probability of depolarizing errors
for each single qubit, the error probability has to decrease
with $O(1/n)$ if large-scale entanglement 
should be maintained \cite{JB2}. 
This strongly suggests that the chromatic index 
or the strength of the interaction has to increase in the order of $n$.
 
For nearest-neighbor interactions in solid state physics
it is not useful to apply our estimations, since there are
considerably tighter bounds (see \cite{BR} paragraph 6.2.1)
But 
for mean-field interactions \cite{DW}, where
 the weighted chromatic index $W(t)$
is already determined
 by the strength of the pair-interactions, we cannot see
any obvious tighter bounds.
Hence physical systems like solid states
with  long-range interactions will present  a typical application 
of our results.  
\end{multicols}

\section*{Acknowledgments}

An important part of the ideas presented above have been developed during
a discussion with my colleague R. Steinwandt.
Part of this work has been supported by the European Union (project Q-ACTA).

\end{document}